# Title: State-dependence of $CO_2$ Forcing and its Implications for Climate Sensitivity

**Authors:** Haozhe He[1]*, Ryan J. Kramer[2,3], Brian J. Soden[1], Nadir Jeevanjee[4]

**Affiliations:**

[1]Rosenstiel School of Marine, Atmospheric and Earth Science, University of Miami, Miami, FL, USA.

[2]Climate and Radiation Laboratory, NASA Goddard Space Flight Center, Greenbelt, MD, USA.

[3]Goddard Earth Science Technology and Research II, University of Maryland at Baltimore County, Baltimore MD, USA.

[4]Geophysical Fluid Dynamics Laboratory, Princeton, NJ, USA.

*Corresponding author. Email: haozhe.he@miami.edu

**Abstract:** When evaluating the effect of $CO_2$ changes on the earth's climate, it is widely assumed that instantaneous radiative forcing from a doubling of a given $CO_2$ concentration ($IRF_{2 \times CO2}$) is constant and that variances in climate sensitivity arise from differences in radiative feedbacks, or a dependence of these feedbacks on the climatological base-state. In this paper, we show that the $IRF_{2 \times CO2}$ is not constant, but also depends on the climatological base-state, increasing by ~25% for every doubling of $CO_2$, and has increased by ~10% since the pre-industrial era primarily due to stratospheric cooling, implying a proportionate increase in climate sensitivity. This base-state dependence also explains about half of the intermodel spread in $IRF_{2 \times CO2}$, a problem that has persisted among climate models for nearly three decades.

**One-Sentence Summary:** Carbon dioxide becomes a more potent greenhouse gas as the climate changes in response to increased carbon dioxide.

**Main Text:** Radiative forcing (RF) refers to a change in net radiative flux at the top-of-atmosphere (TOA) due to an externally-imposed perturbation in the earth's energy balance (*1, 2*), such as anthropogenic activities (e.g., emission of greenhouse gases and aerosols) or natural events (e.g., volcanic eruptions). The earth subsequently warms or cools to counteract the flux perturbation and restore radiative equilibrium. The RF is commonly separated into two parts (*1, 3–5*): instantaneous radiative forcing (IRF), which measures the change in net radiative flux that results only from the change in forcing agents, and rapid adjustments (RAs), which consist of radiative perturbations induced by atmospheric responses to the IRF independent of any change in surface temperature. This study focuses on the IRF, considered the best understood aspect of RF (*6*). For $CO_2$ perturbations, the IRF is responsible for approximately two-thirds of the total RF and is the fundamental driver of RAs (*1, 3–5, 7–11*). However, several previous studies have shown that the IRF from a doubling of $CO_2$ concentration ($IRF_{2 \times CO2}$) varies by ~50% among climate models (*9, 12–14*). Although this spread has persisted for nearly three decades, its underlying cause has never been fully resolved.

Climate sensitivity is formally defined as the change in global-mean surface temperature required to restore radiative equilibrium in response to a doubling of $CO_2$ concentration ($\Delta T_{2 \times CO2}$) and is the most widely used metric to quantify the susceptibility of the climate to an externally forced change; i.e., $\Delta T_{2 \times CO2} = -RF_{2 \times CO2}/\lambda$, where the radiative damping ($\lambda$ in W m$^{-2}$ K$^{-1}$) is the efficiency at which radiative equilibrium is restored per unit change in surface temperature. The radiative damping depends on a number of well and not-so-well understood feedbacks within the climate system, and is widely recognized to both vary between climate models and vary in time as the climatological base-state evolves. However, the intermodel variance in the $RF_{2 \times CO2}$ and its dependence on the base-state are less well recognized. In this study, we demonstrate that the $IRF_{2 \times CO2}$ is not a constant, but also depends on the climatological base-state, as suggested by a recent analytical model (*15*). This state-dependence not only explains about half of the intermodel variance in $IRF_{2 \times CO2}$, but fundamentally reshapes our understanding of climate sensitivity with significant implication for both past and future climate changes.

**Results**

The Coupled Model Intercomparison Projects (CMIP), provide a series of coordinated experiments performed in support of the IPCC assessments in which model simulations are performed by using identical emission scenarios (*16, 17*). However, because determining the IRF requires additional calculations, it is not routinely computed for most experiments. In the first comprehensive RF comparison among climate models, Cess et al. (*12*) found that the $IRF_{2 \times CO2}$ ranged from roughly 3.3 to 4.7 W m$^{-2}$. Subsequent studies with newer generations of models found a similar range (*9, 13*). This spread was thought to mainly arise from intermodel differences in the parameterization of infrared absorption by $CO_2$ (*14*).

Double-call radiative transfer calculations are the most direct method for diagnosing the IRF in model simulations. To produce these specialized online diagnostics, a second call is made to the radiation scheme at each timestep and radiative fluxes are re-calculated with a hypothetical forcing agent perturbation, such as $CO_2$ at some increased concentration. These perturbations are solely used to diagnose the IRF and do not interact with the model simulation. Although only a few online double-call calculations were performed by climate models from CMIP5/6, the available output is particularly useful for investigating the state-dependence of $CO_2$ IRF. To avoid the complicating effects of clouds in masking the IRF (*6, 18*), we further simplify our analysis by limiting it to infrared $CO_2$ forcing at the TOA under clear-sky conditions.

Figure 1A shows the *online* double-call calculations available from CMIP5/6 models for the historical AMIP experiment, which contains the most online double-call calculations of any of the CMIP experiments (12 out of 80 participating models provided calculations for this experiment; Tables S1 and S2). The amip experiment consists of atmosphere-only model simulations that all use identical, time-varying sea surface temperatures observed over the period 1979–2008 as boundary conditions. The online double-calls provided are for 4×CO2; note that $IRF_{4 \times CO2} \approx 2 \times IRF_{2 \times CO2}$ for a given climate state (see Materials and Methods). The results exhibit a large intermodel spread (ranging from ~4 to 8 W m$^{-2}$), consistent with that observed in previous model generations (*14*).

To investigate the extent to which differences in the thermal structure of the climatological base-state can explain the intermodel spread of IRF, we perform *offline* double-call $IRF_{4 \times CO2}$ calculations using original atmospheric profiles from the AMIP models and a single radiative transfer model (SOCRATES; see Materials and Methods). In contrast to the *online* counterparts, the same radiative transfer parameterization is used in all of the offline calculations, so their inter-model spread is only due to differences in the climatological base-states. The strong correlation (*r*=0.82) between the IRFs from the online and offline double-call calculations (Fig. 1B) suggests that more than half of the intermodel variance in $IRF_{4 \times CO2}$ results from differences in climatological base-states, not differences in representing the spectral absorption of CO2. This is consistent with a recent study by Pincus et al. (*18*) who computed IRF from different radiative transfer schemes but using the same climatological base-state and found a much smaller spread in $IRF_{4 \times CO2}$ than in the online double-calls (Fig. 1A). Together, these studies provide compelling evidence to suggest that intermodel differences in the climatological base-state is an important contributor to the spread in CO2 IRF.

The influence of the base-state on CO2 IRF is more clearly illustrated in the coupled model simulations from CMIP6 in which a 1% per year increase is imposed in the atmospheric CO2 concentration (1pctCO2; Fig. 2). Although only 2 models (solid lines in Fig. 2A) submitted online double-call calculations, the results reveal a dramatic growth in $IRF_{4 \times CO2}$ as the climatological base-state evolves. For both models, $IRF_{4 \times CO2}$ increases from ~5 W m$^{-2}$ when $IRF_{4 \times CO2}$ is computed in a pre-industrial climate to ~8 W m$^{-2}$ when it is computed in an elevated-CO2 climate. This challenges the widely held assumption that the $IRF_{2 \times CO2}$ is constant (*19, 20*). To the contrary, it demonstrates that the CO2 IRF is a dynamic quantity that changes substantially as the climate changes.

To verify this result, we perform a series of line-by-line and SOCRATES offline double-call calculations using the full suite of CMIP5/6 coupled simulations under the 1pctCO2 scenario (Fig. 2A, markers). These results both confirm the dramatic increase in $IRF_{4 \times CO2}$ using a much larger ensemble of models and, since the same radiative transfer scheme is used for all offline calculations, indicate that changes in the climatological base-state are responsible for this increase. Note that the climatological base-state here includes the thermal structure as well as the base-state CO2 concentration (*21, 22*), both of which vary with each timestep. However, most of the $IRF_{4 \times CO2}$ increases are due to the evolution of thermal structure, especially for the first doubling of base-state CO2 concentration (Fig. S1).

According to the analytical model of Jeevanjee et al. (*15*), the dependence of CO2 IRF on the climatological base-state can be understood in terms of a dependence on the emission temperature of both stratosphere and troposphere as follows:

$$\mathcal{F} = 2l \ln\left(\frac{q_f}{q_i}\right)\left[\pi B(v_0, \overline{T}_{em}) - \pi B(v_0, T_{strat})\right]$$

where $l$ is the 'spectroscopic decay' parameter of 10.2 cm$^{-1}$, $q_i$ is the initial $CO_2$ concentration, $q_f$ is the final $CO_2$ concentration and $\pi B(v_0, \overline{T}_{em} / T_{strat})$ is the hemispherically integrated Planck function at peak absorption wavenumber of $CO_2$ with either the tropospheric emission temperature or stratospheric emission temperature (see Materials and Methods). The latter refers to the temperature of the upper stratosphere, where unit optical depth is achieved by the peak of the $CO_2$ absorption band, while the former depends on surface temperature and free-troposphere relative humidity. This model has been used to help explain the spatially inhomogeneous distribution of IRF that results from a spatially uniform increase of $CO_2$ (*23*).

As $CO_2$ increases in the 1pctCO2 simulations, the surface temperature warms and the stratosphere cools roughly linearly over time (Figs. 2B and 2C). To assess the relative contributions of these changes in climate to the increase in IRF$_{4\times CO_2}$, we include results from the CMIP6 abrupt-4×CO2 experiment (Fig. 2, dashed lines; only 1 model provided online double-call calculations for this experiment). In contrast to the 1pctCO2 experiment, $CO_2$ is instantly quadrupled in the abrupt-4×CO2 experiment causing the surface to warm rapidly over the first few decades before leveling off. The stratosphere adjusts even more rapidly, equilibrating to a new temperature within the first year.

The contrasting temporal evolution of the climate between these two scenarios is reflected in the IRF$_{4\times CO_2}$. For instance, the IRF$_{4\times CO_2}$ with abrupt-4×CO2 base-state exhibits only a mild increase with global-mean surface warming (Fig. 2), indicating a relatively weak dependence of the $CO_2$ IRF on surface temperature. In contrast, IRF$_{4\times CO_2}$ in the 1pctCO2 experiment exhibits a much larger increase over time, despite having a similar change in global-mean surface temperature. Physically, the $CO_2$ IRF represents a swap of tropospheric emission for stratospheric emission (*15*), and since the temperature change within the stratosphere is much larger than that at surface and within troposphere, the IRF increase closely follows the stratosphere cooling, suggesting a dominant role of stratospheric temperature on the $CO_2$ IRF. As cloud masking has virtually no influence on the stratospheric emission, the dominant role of stratospheric temperature also remains under all-sky conditions.

The state-dependence of $CO_2$ IRF on the surface temperature and stratospheric temperature is also evident in the amip simulations (Fig. 1A). Since these simulations adopt the same sea surface temperature as their boundary conditions, our results imply that differences in stratospheric temperature are largely responsible for the intermodel spread in IRF$_{4\times CO_2}$. To confirm the role of the stratospheric temperature on the IRF spread, we also perform the SOCRATES offline double-call IRF calculations using the same amip base-states and check its correlation with the corresponding air temperature at 10 hPa, which is the highest level of CMIP5 standard pressure-level outputs [and is closest to the level with unit optical depth achieved by the peak of the $CO_2$ absorption band (*15*)]. A high, significant correlation is found between the IRF and stratospheric temperature across both CMIP6 and CMIP5 models (Figs. 1C and S2), highlighting that biases in stratospheric temperature play a dominant role in causing the intermodel spread in $CO_2$ IRF.

The overwhelming role of stratospheric temperature over surface temperature is also reflected in the brief declines for many models in the magnitude of the IRF$_{4\times CO_2}$ at year 1992, following the eruption of Mount Pinatubo (Fig. 1A). For instance, on average across the models there was only a 0.2 K surface temperature decrease but a ~1 K temperature increase at 10 hPa in 1992 compared to 1991.

The analytical model of $CO_2$ IRF by Jeevanjee et al. ([15]) replicates the offline double-call $IRF_{4\times CO2}$ of CMIP6 and CMIP5 with high correlations for abrupt-4×CO2 simulations (Figs. 3A and S3), providing a computationally efficient alternative for investigating the sensitivity of the $CO_2$ IRF to stratospheric temperatures. Since the 10 hPa temperatures cool at a similar rate for all models under 1pctCO2 scenarios from CMIP6 and CMIP5 (Figs. 3B and S4), the temperatures at this level have nearly identical intermodel spread at the beginning and the end of the simulations. This suggests that intermodel spread in the $CO_2$ IRF specifically arises from differences in the initial stratospheric temperatures under pre-industrial conditions. We confirm this with the analytical model, finding it produces the same IRF intermodel spread, highly correlated with the offline double-call calculations, even when the initial, pre-industrial upper stratospheric temperatures are used as input for every timestep instead of the actual, time varying temperature from the corresponding abrupt-4×CO2 simulations (Figs. 3C and S5).

Briefly, our results demonstrate that $CO_2$ IRF increases as the climate changes in response to increased $CO_2$. Online and offline double-call calculations from the CMIP6 historical simulations (Figs. 4A and S6C) indicate that $IRF_{4\times CO2}$ is about 10% larger today than it was in the mid-19[th] century due to the change in base-state, primarily from stratospheric cooling. Thus the $IRF_{2\times CO2}$ is not constant, but varies in time. Since it is the total or "effective" RF that ultimately drives climate change ([1, 3, 24]) and the total RF is the sum of IRF and RAs, it is important to understand the extent to which the RAs may also depend on the base-state.

The most important of the RAs to $CO_2$ forcing is the stratospheric adjustment, which is typically an order of magnitude larger than the tropospheric adjustment ([10, 11]). To investigate the state-dependence of the adjustments, we use atmosphere-only model simulations forced by boundary conditions of preindustrial era (piclim-control) and recent warming decades (amip) along with their corresponding 4×CO2 counterparts (piclim-4×CO2 and amip-4×CO2; see Materials and Methods). The amip simulation not only has a higher prescribed $CO_2$ concentration than that of piclim-control simulation, but also has cooler stratosphere temperature, allowing us to quantify the magnitude of RAs under two different base-states. Figure S6 compares the total RF, stratospheric adjusted RF, IRF, and stratospheric adjustments from the $CO_2$ quadrupling for the two different base-states. The amip simulations exhibit a larger IRF (0.38 W m$^{-2}$) compared to that obtained under preindustrial conditions due to the cooler stratosphere. A nearly identical increase is seen in both the stratospheric adjusted RF (Fig. S6B; 0.34 W m$^{-2}$) and total RF (Fig. S6A; 0.29 W m$^{-2}$), because the stratospheric adjustment is nearly identical between the two sets of experiments (Fig. S6D; –0.03 W m$^{-2}$). Note the abovementioned ensemble-mean forcing differences are also corroborated by differences shown for individual models. Even though radiative transfer scheme differences exert noticeable impacts on the total RF and stratospheric adjustment, a high, significant correlation is also found between the IRF and either total RF (Fig. S7A) or stratospheric adjusted RF (Fig. S7B). Thus, the base-state dependence of the IRF propagates through to a nearly identical dependence in the total RF (Figs. S6 & S7) and thus on climate sensitivity.

Changes in climate sensitivity can therefore arise from both changes in climate feedbacks as well as changes in IRF. More generally, these results indicates that, despite the logarithmic dependence of $CO_2$ absorption, the climate becomes increasingly sensitive to a doubling of $CO_2$, as the base-state $CO_2$ concentration increases and stratosphere cools correspondingly. Since the $IRF_{2\times CO2}$ increases by ~25% for each doubling of base-state $CO_2$ concentration (the $IRF_{2\times CO2}$ increases by 24% and 29% for the first and second doubling of base-state $CO_2$ concentration, respectively; Fig. 2A) and as the IRF accounts for roughly two-thirds of the total RF from $CO_2$ ([1, 9–11]), this implies that $\Delta T_{2\times CO2}$ increases by ~15–20% for each doubling of $CO_2$ just due to changes in IRF.

**Potential Climate implications**

Since the upper stratospheric temperature plays a dominant role in determining the magnitude of the $CO_2$ IRF, including $IRF_{2\times CO_2}$ as well as $CO_2$ greenhouse effects (hereinafter referred to as "climatological $CO_2$ IRF"; Fig. S8), any changes in atmospheric composition that perturbs stratospheric temperature could subsequently impact the climate. Considering the recent example of polar ozone depletion (*25–27*), as solar absorption by ozone strongly influences the temperature structure within stratosphere (*28*), the ozone depletion has led to strong cooling within stratosphere and further amplification of climatological $CO_2$ IRF as our results suggest (Fig. 4A). Note that although the stratospheric ozone loss mainly occurs in the lower stratosphere (*29, 30*), the associated cooling also contributes to a decline in infrared emission from the lower to upper stratosphere, and thus a strengthening of the climatological $CO_2$ IRF at the TOA.

Here, we diagnose the surface warming caused specifically by the amplification of the climatological $CO_2$ IRF resulting from ozone depletion. This warming is quantified as the nonlinear contribution (see Materials and Methods) to the ensemble- and time-mean, total surface temperature anomalies in fully coupled simulations of 1985–2014 using realistic forcing, a period when stratospheric ozone remained at dangerously low levels (Fig. 4B). As expected, most of the indirect surface warming effect occurs around the poles, where the local stratosphere has the strongest cooling, although some heat transport may also be playing a role (*31*). This supports the premise that any forcing agent changes that perturb the stratospheric temperature could also impact the climate by modulating the $CO_2$ IRF at the TOA, even without changing $CO_2$ amount.

Our findings may also help to better understand past climate events, such as the end-Devonian mass extinction and the Paleoproterozoic "snowball earth" conditions, occurred following similar but considerably stronger perturbations, i.e., a dramatic drop in stratosphere ozone (*32*) and the inevitable development of an ozone layer (*33, 34*), respectively. Meanwhile, this base-state dependence of the $CO_2$ IRF may have implications for how other related metrics are defined, such as global warming potential and efficacy of non-$CO_2$ forcing (*8, 24*), since they are quantified relative to the radiative effects of a $CO_2$ perturbation. These metrics are often used in policy discussions, so it will be particularly important to determine if they must be re-defined with a better consideration of the non-constant $CO_2$ IRF effects.

Additionally, our results could also have implications for geoengineering and climate change mitigation (*35*). Taking 1992 - the year following the 1991 eruption of Mount Pinatubo - as an example, the injected volcanic aerosols within stratosphere not only cooled the surface by reflecting more solar radiation back to the space but also warmed the stratosphere by increasing the atmospheric absorption of sunlight in the stratosphere (*36, 37*). The resulting stratospheric warming also weakened the $CO_2$ IRF (Figs. 1A and 4A) and mitigated the warming efficacy of $CO_2$. As most geoengineering approaches involving stratospheric aerosol injection employ reflective aerosols [e.g., sulfate (*38*)], alternative approaches that involve the use of more absorbing aerosols (e.g., black carbon) may warrant consideration, as it could more effectively reduce $CO_2$ IRF by further warming the upper stratosphere (*39, 40*).

Lastly, we note that the model simulations of stratospheric temperature can be easily constrained with observations. Across multiple sets of observations and reanalyses (see Materials and Methods), the global- and annual-mean 10 hPa air temperature has an uncertainty range of 227.2 to 228.3 K at year 2020. This ~1.1 K difference in base-state would translate to only a ~0.10 (0.11) W m$^{-2}$ $IRF_{4\times CO_2}$ uncertainty for CMIP6 (CMIP5) models (Figs. 1C and S2). This highlights the importance of accurately representing the stratosphere when projecting future $CO_2$-induced climate change and the potential to better constrain model projections using observations, further emphasizing the importance of observations in Earth's middle and upper atmosphere (*41*).

**Acknowledgments:** We thank Drs. Adriana Sima, Chris Smith and Pierre Nabat for clarifying CMIP standard online double-call methods and Dr. Jacob Seeley for insightful discussions at the initial stage of this work. We also thank Drs. David Paynter and Pu Lin for their helpful comments.

**Funding:**

NOAA Award NA18OAR4310269 (HH, RJK, BJS)

NOAA Award NA21OAR4310351 (HH, BJS)

NASA Science of Terra, Aqua and Suomi-NPP grant 80NSSC21K1968 (RJK)

**Author contributions:**

Conceptualization: HH, RJK, BJS, NJ

Methodology: HH, RJK, BJS, NJ

Investigation: HH, RJK, BJS, NJ

Visualization: HH

Funding acquisition: RJK, BJS, NJ

Project administration: BJS

Supervision: BJS

Writing – original draft: HH, RJK, BJS

Writing – review & editing: HH, RJK, BJS, NJ

**Competing interests:** Authors declare that they have no competing interests.

**Data and materials availability:** The CMIP6 data are available at https://esgf-node.llnl.gov/search/cmip6/ while CMIP5 data are available at https://esgf-node.llnl.gov/projects/cmip5/. The CMIP6/5 models used in this work are listed in Tables S1–S4 in the Supplementary Materials. The AIRS temperature observations, Aqua IR-only, SNPP, and NOAA-20 products produced using the CLIMCAPS algorithm and the MERRA-2 reanalysis data are available at https://disc.gsfc.nasa.gov/datasets/. The ERA5 reanalysis data are available at https://cds.climate.copernicus.eu/cdsapp#!/dataset/reanalysis-era5-pressure-levels-monthly-means?tab=overview. The NCEP-DOE Reanalysis 2 are available at https://psl.noaa.gov/data/gridded/data.ncep.reanalysis2.pressure.html. The benchmark radiative forcing values are obtained online (https://github.com/RobertPincus/rfmip-benchmark-paper-figures). SOCRATES is available from https://code.metoffice.gov.uk/trac/socrates but requires a free account from the UK Met Office to access the website. ARTS is available at https://www.radiativetransfer.org/getarts/ while PyRADS is available at https://github.com/danielkoll/PyRADS. Codes to produce the paper are available from the corresponding author upon request.

## Supplementary Materials

Materials and Methods

Figs. S1 to S8

Tables S1 to S4

References (*42–63*)

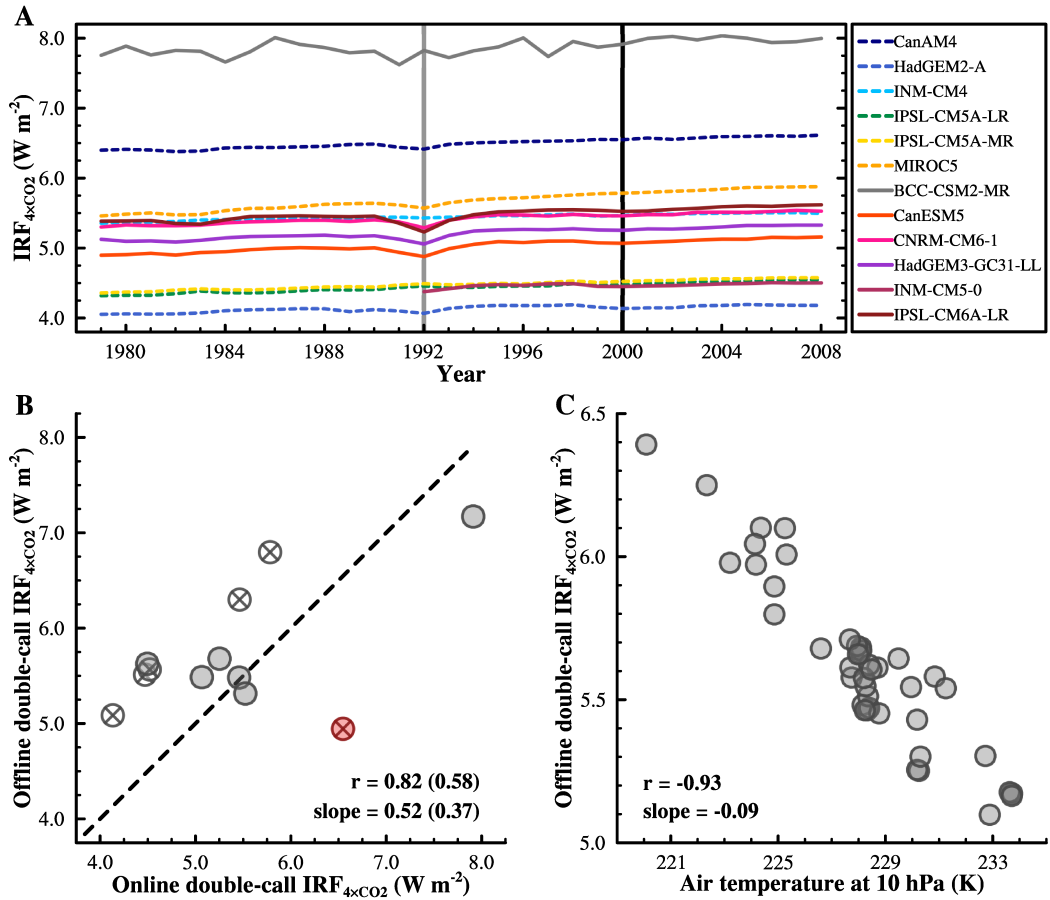

**Fig. 1. The intermodel spread in IRF$_{4\times CO_2}$ and its causes.** (**A**) Time series of all available online double-call IRF$_{4\times CO_2}$ with base-state from amip experiments for CMIP5/6 models. The black vertical reference line highlights the IRF$_{4\times CO_2}$ values used in (B), while the gray one accentuates the brief declines in the magnitude of the IRF$_{4\times CO_2}$ at year 1992, following the eruption of Mount Pinatubo. (**B**) A comparison of the IRF$_{4\times CO_2}$ at year 2000 from the online and offline double-call calculations. The gray filled circles represent models from CMIP6 while the open circles with cross inside show models from CMIP5. The red filled circle with cross inside highlights the outlier model (i.e., CanAM4). Since the vertical IRF profile of CanAM4 shows an increase with height within stratosphere [Fig. 3 of Chung and Soden (*9*)], it differs from the common expectation based on the negative lapse rate within stratosphere. It is reasonable to exclude results of the CanAM4 from the spread contribution analyses. The values in front of (in) brackets shown in (B) are values calculated without (with) the outlier model CanAM4. (**C**) A scatterplot of global- and annual-mean air temperature at 10 hPa of each model at year 2000 of amip experiment versus its corresponding offline double-call IRF$_{4\times CO_2}$.

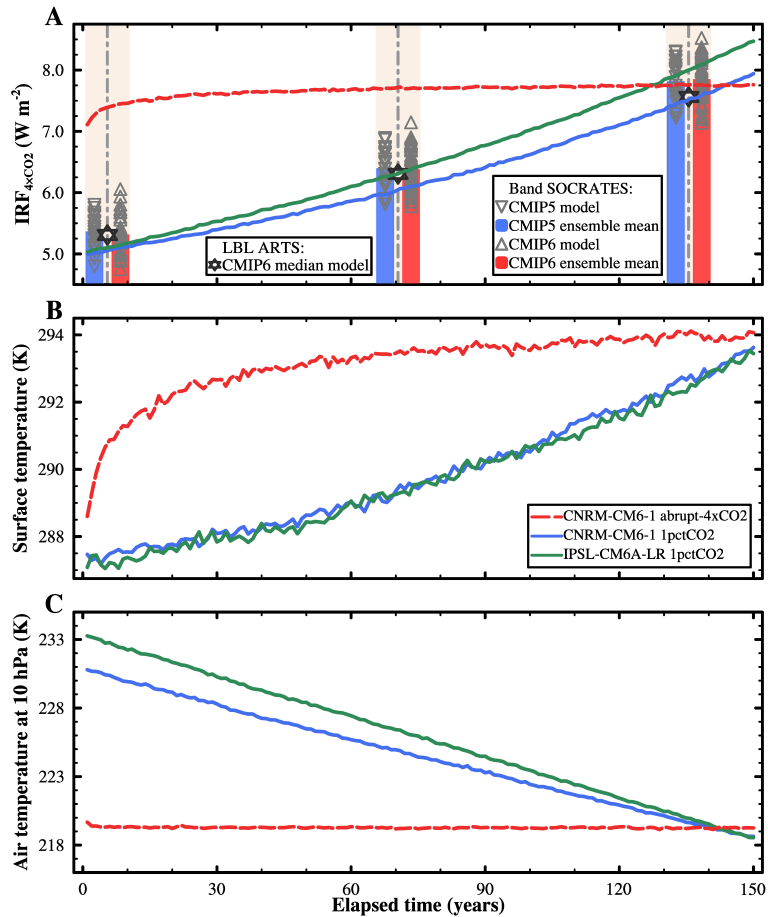

**Fig. 2. The CO₂ IRF increases as the surface warms and stratosphere cools.** Time series of global- and annual-mean (**A**) online double-call $IRF_{4\times CO_2}$, (**B**) surface temperature, and (**C**) air temperature at 10 hPa from models CNRM-CM6-1 and IPSL-CM6A-LR. Three highlighted time slices in (A) are years 1–10, 66–75 and 131–140. Overlaid gray triangles represent the global- and time-mean SOCRATES offline double-call $IRF_{4\times CO_2}$ with corresponding atmospheric profiles of 1pctCO2 simulations from CMIP5/6 models. The black stars show the global-mean ARTS offline double-call $IRF_{4\times CO_2}$ with time-mean atmospheric profiles from the CMIP6 model, which has the median SOCRATES double-call $IRF_{4\times CO_2}$ value. Similar results from another line-by-line model (PyRADs) are shown in Fig. S1.

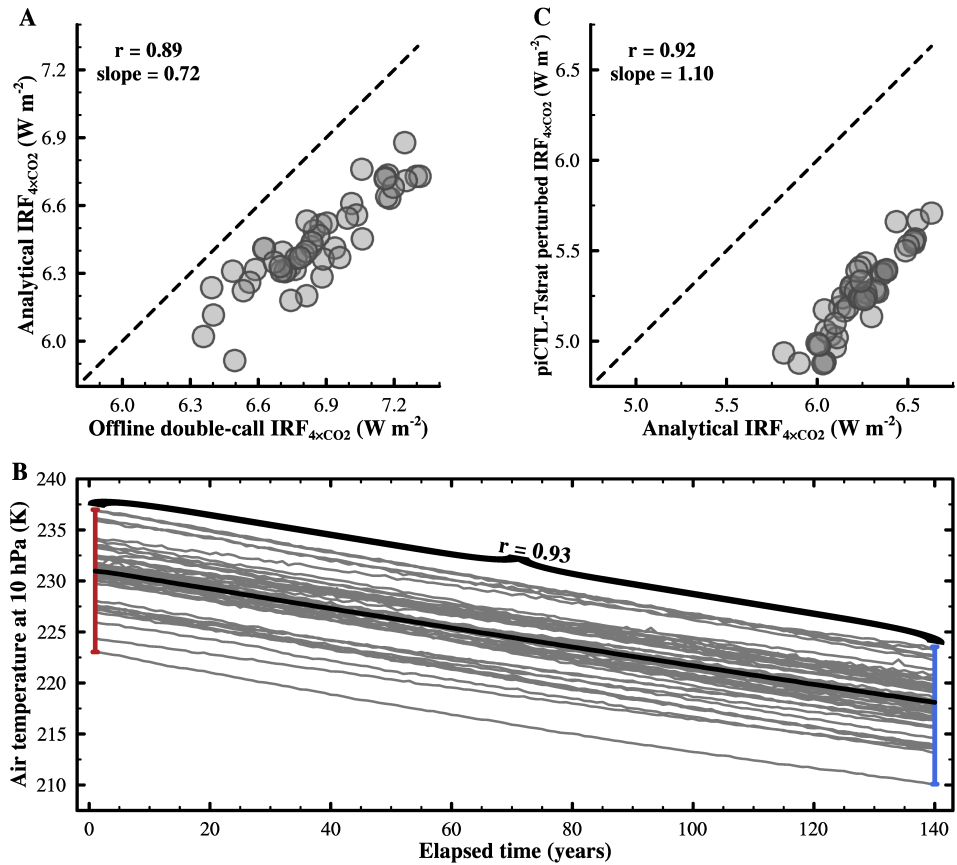

**Fig. 3. Differences in initial stratospheric temperatures across models explain roughly half of the intermodel spread in IRF$_{4\times CO_2}$, as shown using abrupt-4×CO2 experiments.** (**A**) A comparison of global- and time-mean IRF$_{4\times CO_2}$ in year 121-140 from the offline double-call and analytical model calculations with base-state from abrupt-4×CO2 experiments. The correlation between global- and time-mean IRF$_{4\times CO_2}$ in every 10 of 150-year experiments from the offline double-call and the analytical model calculations has a range from 0.88 to 0.89. (**B**) Time series of global- and annual-mean 10 hPa air temperature under 1pctCO2 scenario from CMIP6 models. Each gray line in (B) represents the 10 hPa temperature evolution of a model while the thick black line shows the multi-model ensemble mean. The bracket in (B) highlights the correlation between 10 hPa air temperature at years 1 and 140. (**C**) A comparison of the global- and time-mean original analytical IRF$_{4\times CO_2}$ in year 2-11 and that obtained with perturbed stratospheric emission temperature from piControl runs (piCTL-Tstrat). The correlation between the global- and time-mean IRF$_{4\times CO_2}$ from the original and piCTL-Tstrat perturbed calculations has a range from 0.90 to 0.92.

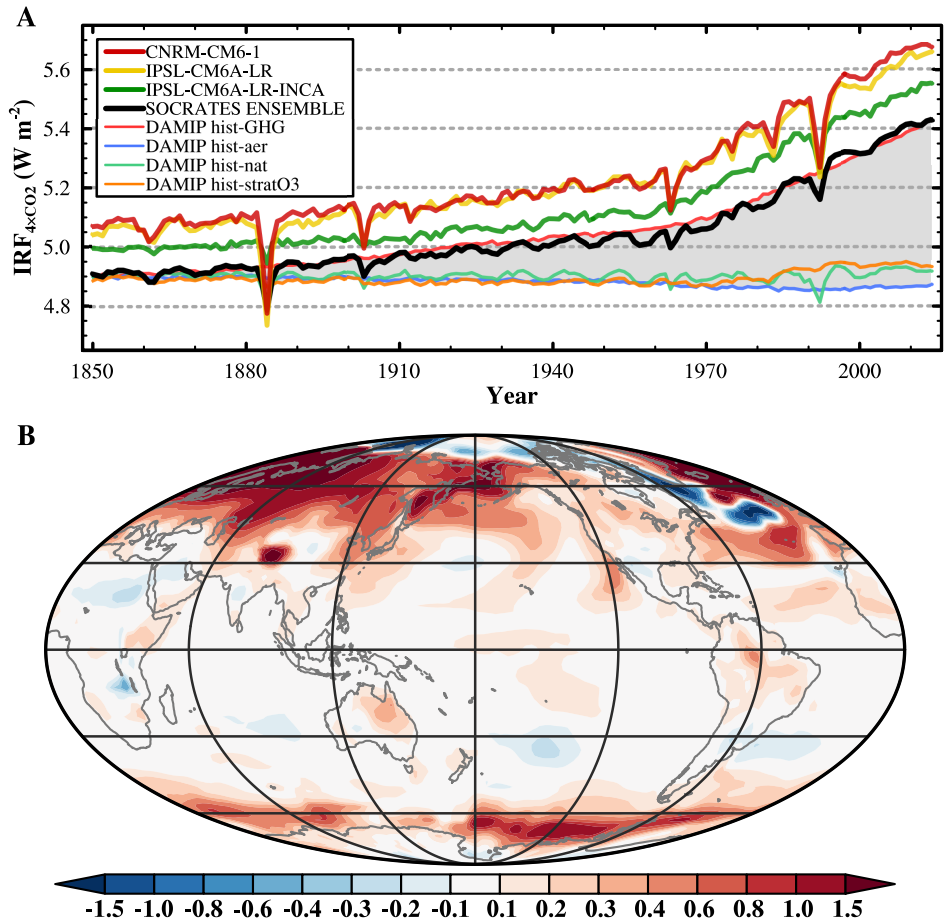

**Fig. 4. Any forcing agent changes that perturb the stratospheric temperature can further impact the climate by modulating the radiative forcing by CO₂ without changing CO₂ amount. (A)** Time series of three available online double-call $IRF_{4\times CO_2}$ with base-state from CMIP6 historical simulations and multi-model ensemble mean of corresponding offline double-call $IRF_{4\times CO_2}$ for five CMIP6 models with all experiments available shown in shading highlighted thin lines. **(B)** The ensemble-mean map of the indirect surface warming effect of ozone depletion during period 1985-2014.